\documentclass[aps,prb,twocolumn,superscriptaddress]{revtex4-1}
\usepackage{hyperref}
\usepackage{dsfont}
\usepackage{amsmath}
\usepackage{amssymb}
\usepackage{graphicx}
\usepackage{color}

\begin{document}

\title{Highlighting superconductivity in an insulator}

\author{A.~D. Mirlin}
\affiliation{Institut f\"ur Nanotechnologie, Karlsruhe Institute of Technology, 76021 Karlsruhe, Germany}
\affiliation{Institut f\"ur Theorie der Kondensierten Materie, Karlsruhe Institute of Technology, 76128 Karlsruhe, Germany} 
\affiliation{Petersburg Nuclear Physics Institute, 188350 St. Petersburg, Russia}
\affiliation{Landau Institute for Theoretical Physics, 119334 Moscow, Russia}
\author{I.~V. Protopopov}
\affiliation{Department of Theoretical Physics, University of Geneva, 1211 Geneva, Switzerland  }
\affiliation{Landau Institute for Theoretical Physics, 119334 Moscow, Russia}

\date{\today}

\pacs{}

\maketitle



{\bf Superconductor-insulator transition  is a fascinating quantum phenomenon that reveals a competition between phase order and charge localization. 
Microwave spectroscopy provides a novel promising approach to its controllable investigation in 1D Josephson arrays.}

\vspace*{0.3cm}

Superconductivity and Anderson localization are two quantum phenomena playing an outstanding role in the condensed matter physics. They are in a certain sense antagonists: while a superconductor has zero resistivity, an Anderson insulator  is characterized by an infinite resistivity in the zero-temperature limit. Remarkably, one-dimensional (1D) and two-dimensional (2D) systems undergo a direct quantum phase transition from one of these extremes to the other---the superconductor-insulator transition (SIT). Despite considerable progress in the past years, experimental analysis of the physics of these quantum phase transitions remains a great challenge. Now, writing in {\it Nature Physics}, Roman Kuzmin and colleagues \cite{kuzmin19} make a major step in this direction. Specifically, they develop a new tool to investigation of the vicinity of SIT in a Josephson junction array based on microwave spectroscopy of collective bosonic excitations---plasmons---characteristic for 1D superconductor.  

The SIT in 1D systems is driven by an interplay of disorder with quantum fluctuations of the superconducting order parameter---quantum phase slips (QPS). A type of disorder that is particularly important for Josephson-junction arrays is random stray charges. The transition is of Berezinskii-Kosterlitz-Thouless character, with the fixed point that controls properties near the transition in the renormalization-group analysis being located at zero value of fugacity governing the QPS  \cite{bradley84,giamarchi88,bard17}.  This permits detailed theoretical predictions concerning properties of the systems around the transition. The situation in 1D is in this sense more favorable than in 2D where the SIT fixed point is located at strong coupling, making the theory analysis of the SIT vicinity much more difficult. 

On the experimental side, Josephson-junction chains have been shown to be a suitable platform for 1D SIT \cite{chow98}. On the other hand, original experiments on transport properties of chains had substantial difficulties with location of SIT. The source of these problems is not quite clear; it is possible that the results were affected by some kind of noise related to the environment.  Recently, a progress was made in a controllable experimental study of transport in the insulating phase \cite{cedergren17}, providing a hope that such experiments will come closer to SIT in near future. 

Kuzmin {\it et al.} now choose a very different approach to this challenging problem. They couple a double chain of Josephson junctions (consisting of as much as 33,000 junctions) to an antenna and investigate positions and width of resonances in the microwave frequency range. This spectroscopic approaches has several important advantages. First, the frequency serves as a new knob that allows one to move effectively from superconductor to insulator. Second, no contacts have to be attached, which is favorable from the point of view of insulation of the system from the environment. The double-chain architecture has allowed Kuzmin and coauthors to focus on the antisymmetric mode that is weakly susceptible to the capacitance to the ground, which is helpful for controlling parameters of the system.

Two parameters that are of central importance for the SIT in this system are the Josephson energy $E_J$ and the inter-chain charging energy $E_0 = (2e)^2/C_0$ where $C_0$ is the capacitance between the corresponding grains of two chains. The dynamics of the superconducting phase can be described by the Luttinger-liquid theory with an interaction constant\cite{bard18} $K= \pi \sqrt{E_J/2E_0}$. The SIT fixed point corresponds to the critical value $K_c=3/2$. The samples studied by Kuzmin et al  have values of $K$ a few times smaller than this, which means that they are on the insulating side of SIT (Fig. 1). Nevertheless, spectroscopy reveals a clear superconducting feature: narrow, nearly equidistant resonances building a branch of bosonic excitations whose dispersion corresponds to the phase fluctuation mode characteristic for a 1D superconductor. So, although system is actually an insulator, it shows its underlying superconducting nature when probed at finite (and not too low) frequency $\omega$. How can one then see that the system is actually an insulator? To this end, Kuzmin et al measure the width of resonances and find that it becomes larger when $\omega$ is lowered. This allows them to observe decay processes induced by QPS \cite{bard18} whose proliferation is a hallmark of the insulator. Another experimental indication of the insulating character of the system at low energies is increasing fluctuations in positions of resonances with lowering $\omega$. 

\begin{figure}
  \centering
  \hspace*{-.05\columnwidth}
    \includegraphics[width=.52\columnwidth]{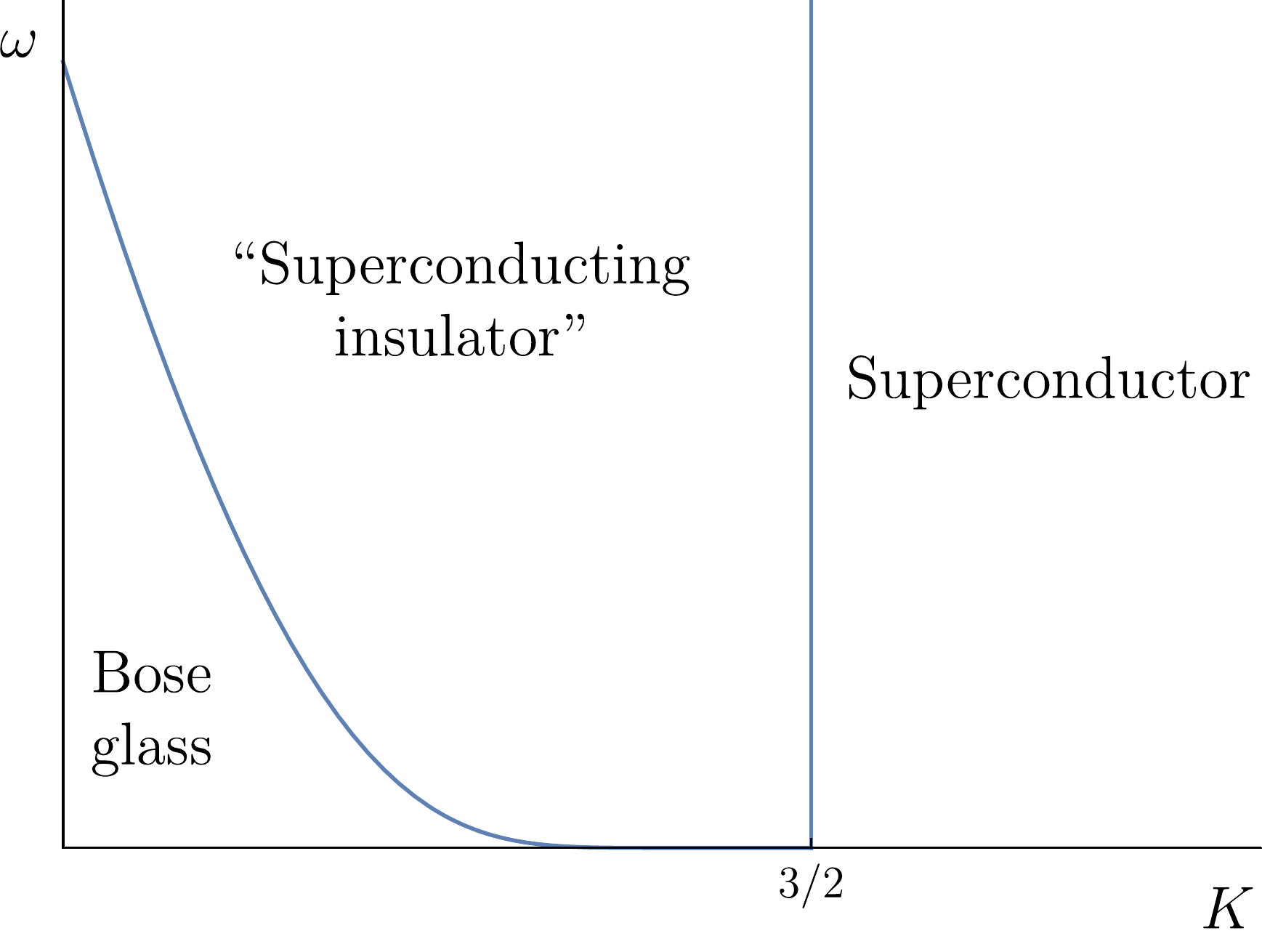} \hspace*{.03\columnwidth}
  \includegraphics[width=.42\columnwidth]{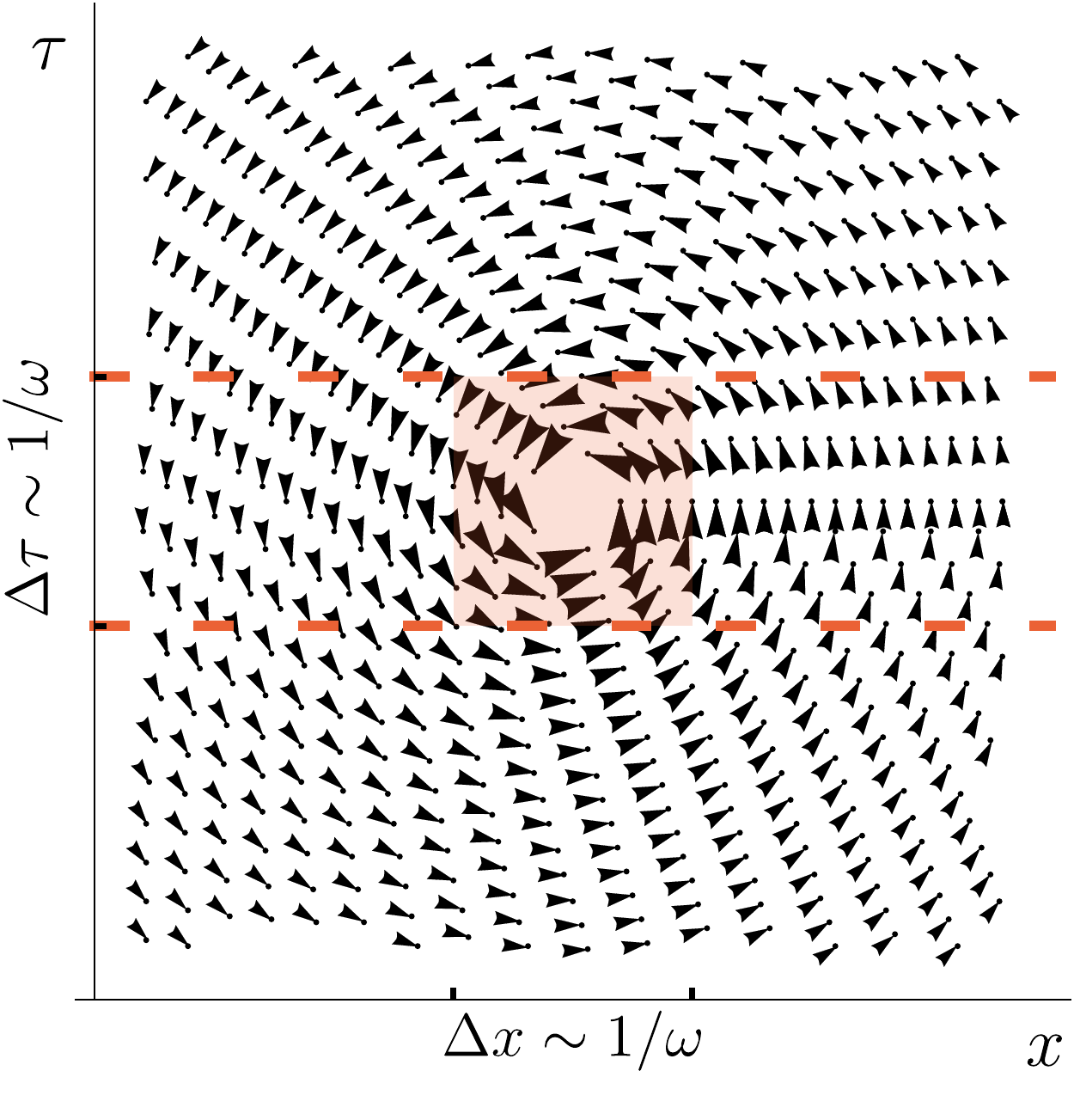}
  \caption{{\it Left:} Schematic phase diagram in the plane spanned by the Luttinger-liquid constant $K$ and the frequency $\omega$.  The SIT at $K = K_c = 3/2$ characterizes the infrared limit, $\omega \to 0$. Kuzmin et al explore the insulating side of the transition at elevated frequencies (``superconducting insulator''). With lowering $\omega$, a crossover to the insulating (``Bose glass'') regime is expected. {\it Right:}  QPS responsible for destruction of the superconducting phase and transition to insulator.  Arrows show the gradient of the superconducting phase, as a function of the coordinate $x$ and (imaginary) time $\tau$. At larger $\omega$, the effective size of the QPS shrinks (shown by red color), which reduces the effect of QPS and thus reveals superconducting properties of the system.}
  \label{fig1}
\end{figure}

The work by Kuzmin and colleagues opens broad prospects for experimental investigations and potential applications of 1D Josephson arrays near SIT. Let us briefly mention some of them. First, a highly challenging goal would be to characterize quantitatively  the frequency dependence of the resonance width and in this way to study experimentally the scaling at the SIT.  Second, it would be very interesting to extend the experimental spectroscopic study to lower frequencies where the system is expected to fully develop it insulating character, becoming a so-called Bose glass \cite{wu18,houzet19}. As a related task, one could look for experimental manifestation of strong-disorder SIT transition that has been proposed theoretically \cite{doggen17}. A further important direction is the investigation of emergent strong-coupling electrodynamics in such devices. Indeed, the Josephson chain can be viewed as a transmission line with a wave impedance $Z = (h/8\pi e^2) \sqrt{2 E_0/E_J}$ related to Luttinger-liquid constant via $Z = (1/8K)h/e^2$. The values of $Z$ near the SIT, as studied by Kuzmin et al, are of the order of 10\:k$\Omega$. This should be compared to the vacuum impedance $Z_0 \simeq 377\:\Omega$. The latter value is connected to the fine structure constant $\alpha \simeq 1/137$ via $\alpha = Z_0 (e^2/2h)$. Via the same token, the large values of $Z$ in the Josephson chain near SIT correspond to effective values of $\alpha$ close to unity. The effective strong coupling quantum electrodynamics can be revealed by coupling an artificial ``atom'' (two-level system) to the electromagnetic field of the Josephson transmission line \cite{kuzmin19a}. Finally, Josephson chains near the SIT can find application as superinductors that are considered as important building blocks for prospective quantum information superconducting devices.


\begin{thebibliography}{99}

\bibitem{kuzmin19}
R. Kuzmin, R. Mencia, N. Grabon, N. Mehta, Y.-H. Lin, and V. E. Manucharyan,
``Quantum electrodynamics of a superconductor-insulator phase transition'', Nature Physics XXXX (2019). 

\bibitem{bradley84} R. M. Bradley and S. Doniach, Phys. Rev. B 30, 1138 (1984). 

\bibitem{giamarchi88}
T. Giamarchi and H. J. Schulz, Phys. Rev. B {\bf 37}, 325 (1988).

\bibitem{bard17}
M. Bard, I. V. Protopopov, I. V. Gornyi, A. Shnirman, and
A. D. Mirlin, Phys. Rev. B {\bf 96}, 064514 (2017).

\bibitem{chow98}
E. Chow, P. Delsing, and D. B. Haviland, Phys. Rev. Lett. {\bf 81},
204 (1998).

\bibitem{cedergren17}
K. Cedergren, R. Ackroyd, S. Kafanov, N. Vogt, and A. Shnirman,
and T. Duty, Phys. Rev. Lett. {\bf 119}, 167701 (2017).

\bibitem{bard18}
M. Bard, I. V. Protopopov, and A. D. Mirlin,
Phys. Rev.  B {\bf 98}, 224513 (2018).

\bibitem{wu18}
H.-K. Wu and J.D. Sau, arXiv:1811.07941.

\bibitem{houzet19}
M. Houzet and L.I. Glazman,  arXiv:1901.01515.

\bibitem{kuzmin19a}
R. Kuzmin, N. Mehta, N. Grabon, R. Mencia, and V. E. Manucharyan, npj Quantum Information  5:20 (2019).

\bibitem{doggen17}
E. V. H. Doggen, G. Lemari\'e, S. Capponi, and N. Laflorencie,
Phys. Rev. B {\bf 96}, 180202(R) (2017).



\end{thebibliography}
\end{document}